\documentclass[%
aip,
 amsmath,amssymb,
 reprint,%
]{revtex4-1}
\usepackage{graphicx}
\usepackage{dcolumn}
\usepackage{bm}

\usepackage[utf8]{inputenc}
\usepackage[T1]{fontenc}
\usepackage{mathptmx}
\usepackage{etoolbox}
\renewcommand{\selectlanguage}[1]{}
\makeatletter
\def\@email#1#2{%
 \endgroup
 \patchcmd{\titleblock@produce}
  {\frontmatter@RRAPformat}
  {\frontmatter@RRAPformat{\produce@RRAP{*#1\href{mailto:#2}{#2}}}\frontmatter@RRAPformat}
  {}{}
}%
\makeatother

\usepackage{hyperref}

\begin{document}

\preprint{AIP/123-QED}

\title[Dielectric microwave resonator with large optical apertures for spin-based quantum devices]{
Dielectric microwave resonator with large optical apertures for spin-based quantum devices}

\author{Tatsuki Hamamoto}
 \altaffiliation{Experimental Quantum Information Physics Unit, Okinawa Institute of Science and Technology Graduate University, Onna, Okinawa 904-0495, Japan}
\author{Amit Bhunia}%
 \altaffiliation{The Science and Technology Group, Okinawa Institute of Science and Technology Graduate University, Onna, Okinawa 904-0495, Japan}
\author{Rupak Kumar Bhattacharya}%
 \altaffiliation{The Science and Technology Group, Okinawa Institute of Science and Technology Graduate University, Onna, Okinawa 904-0495, Japan}
\author{Hiroki Takahashi}%
 \altaffiliation{Experimental Quantum Information Physics Unit, Okinawa Institute of Science and Technology Graduate University, Onna, Okinawa 904-0495, Japan}
\author{Yuimaru Kubo}
 \altaffiliation{The Science and Technology Group, Okinawa Institute of Science and Technology Graduate University, Onna, Okinawa 904-0495, Japan}
 \email{yuimaru.kubo@oist.jp}
\date{\today}

\begin{abstract}
Towards a spin-based quantum microwave-optical photon transducer, we demonstrate a low-loss dielectric microwave resonator with an internal quality factor of $2.30\times10^4$ while accommodating optical apertures with a diameter of $8\, \mathrm{mm}$. 
The two seemingly conflicting requirements, high quality factor and large optical apertures, are satisfied thanks to the large dielectric constant of rutile ($\mathrm{TiO_2}$).
The quality factor is limited by radiation loss, and we confirmed by numerical simulation that this dielectric resonator can achieve a quality factor exceeding $10^6$ by extending the height of the resonator enclosure. 
Using this resonator, we performed both continuous-wave (cw) and pulse electron spin resonance (ESR) spectroscopy on 2,2-diphenyl-1-picrylhydrazyl (DPPH) crystalline powder and P1 centers in a diamond crystal in a dilution refrigerator. 
The cw ESR spectroscopy demonstrated high-cooperativity and strong spin-resonator coupling with the DPPH and P1 centers respectively, while the pulse ESR spectroscopy successfully measured longitudinal and transverse relaxation times.

\end{abstract}

\maketitle



A reliable quantum network requires the connection of quantum computers through robust flying qubits.\cite{wehner_quantum_2018,reiserer_colloquium_2022,ruf_quantum_2021,awschalom_development_2021}
Optical photons are ideal candidates for this task at ambient temperatures. 
On the other hand, in terms of scalability, the most advanced quantum computing platform consists of superconducting circuits,\cite{arute_quantum_2019,krantz_quantum_2019,kjaergaard_superconducting_2020}
where photons operate at microwave frequencies inside a dilution refrigerator at millikelvin temperatures. 
This gap necessitates a quantum wavelength converter, called a quantum transducer, to bridge optical and microwave photons.\cite{wehner_quantum_2018,reiserer_colloquium_2022,awschalom_development_2021} 

In this perspective, several hybrid approaches have been conceived \cite{stannigel_optomechanical_2010,rabl_quantum_2010,tsang_cavity_2010,safavi-naeini_proposal_2011,wang_using_2012,hisatomi_bidirectional_2016,javerzac-galy_-chip_2016,zhong_proposal_2020} and explored.\cite{higginbotham_harnessing_2018,mohammad_mirhosseini_superconducting_2020,jiang_optically_2022,forsch_microwave--optics_2020,hease_bidirectional_2020,xu_bidirectional_2021,okada_superconducting_2021}
In particular,  there have been significant advances in opto-mechanical \cite{higginbotham_harnessing_2018,mohammad_mirhosseini_superconducting_2020,jiang_optically_2022,forsch_microwave--optics_2020} and electro-optical \cite{hease_bidirectional_2020,xu_bidirectional_2021,okada_superconducting_2021} devices over the past decade. 
Impurity spins in solid crystals are also of interest as a promising system for mediating microwave and optical photons for superconducting quantum circuits.\cite{williamson_magneto-optic_2014,barnett_theory_2020,rochman_microwave--optical_2023,fernandez-gonzalvo_cavity-enhanced_2019}
Additionally, a spin ensemble has potential as a multi-mode quantum random access memory (RAM),\cite{julsgaard_quantum_2013,afzelius_proposal_2013,grezes_multimode_2014,wu_storage_2010} offering a pathway to reinforce the robustness of quantum repeaters or computers at each network node.\cite{wehner_quantum_2018,reiserer_colloquium_2022,ruf_quantum_2021} 

In the ``spin ensemble-based'' quantum transduction, achieving an efficient mode match between a microwave resonator and an optical cavity is crucial. \cite{williamson_magneto-optic_2014}
Loop-gap microwave resonators have emerged as a promising candidate to satisfy this prerequisite.\cite{williamson_magneto-optic_2014,fernandez-gonzalvo_cavity-enhanced_2019,ball_loop-gap_2018}
However, it is still not obvious if their internal (unloaded) quality factors are sufficient for high-efficiency transduction, primarily due to the dissipation caused by the finite conductivity of normal metals. 
Using superconductors could mitigate this issue,\cite{rochman_microwave--optical_2023,chen_coupling_2016} although this brings another limitation on the static magnetic fields to the spins because of the Meissner effect. 
Moreover, the necessity of optical apertures for free-space optical beams or fibers leads to radiation losses; resulting in a reduction of the quality factor.  

Dielectric microwave resonator appears to be a promising approach for achieving high internal quality factors.\cite{probst_three-dimensional_2014} 
In this work, we develop a rutile ($\mathrm{TiO_2}$)-based dielectric microwave resonator incorporating optical apertures with diameters of 8 mm, which is larger than the diameter of the sample space, on the top and bottom of the resonator enclosure. 
Through the resonator, we measured an ensemble of free radicals in 2,2-Diphenyl-1-picrylhydrazyl (DPPH) and an ensemble of nitrogen impurities (P1 centers) in a diamond crystal as test samples. 
Despite such sizable optical apertures, the internal quality factor of $Q_{\mathrm{int}} \sim 10^4$ is achieved at both 4 K and 10 mK, enabling the system to reach the high-cooperativity and strong coupling regimes with DPPH and P1 centers respectively. 
In addition, the numerical simulation result indicates that the microwave ac magnetic field is homogeneous ($\approx 95 \%$) over the sample space, with about $37 \%$ of the magnetic energy confined inside the rutile. Details of the numerical simulation are discussed in the supplementary material.
Therefore, our resonator design is advantageous for spin-based quantum devices such as quantum transducers.\cite{williamson_magneto-optic_2014,barnett_theory_2020} 

\begin{figure}
    \includegraphics[width=\hsize]{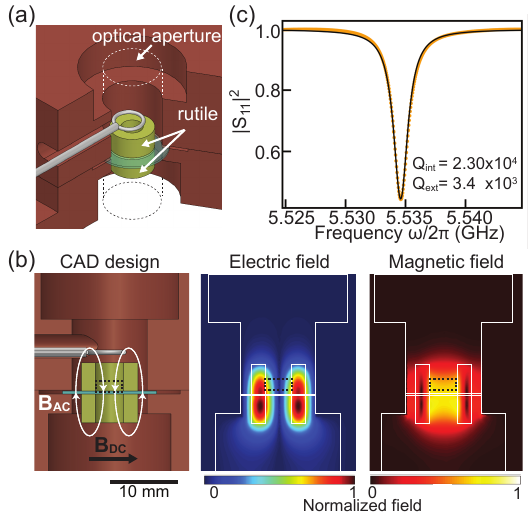}
    \caption{
    The dielectric microwave resonator. 
    (a) A cutaway drawing showing a hollow-cylindrical dielectric crystal (rutile $\mathrm{TiO_2}$, yellow) inside an oxygen-free copper enclosure (brown). A sapphire substrate (light blue) is inserted in-between the two rutile pieces to place a sample. 
    The resonator is inductively coupled to the measurement line through a loop antenna (gray).
    The optical aperture with an 8 mm diameter is indicated by dashed lines.
    (b) Numerically simulated distributions of the electric and magnetic $\mathrm{TE_{01\delta}}$ mode (fundamental mode). The static magnetic field $\mathbf{B_{DC}}$ is applied horizontally, and the microwave magnetic field $\mathbf{B_{AC}}$ passes vertically through the inner hole. The magnetic field penetrates the sample space (dashed black box) vertically, while the electric field is confined in the dielectrics.
    (c) A reflection spectrum $|S_{11} (\omega)|$ measured at 10 mK.
    Orange dots and a black line indicate the experimental data and the fitted reflection spectrum, respectively.}
    \label{fig:1_dielectricResonator} 
\end{figure}

Figure \ref{fig:1_dielectricResonator} summarizes the dielectric resonator. 
A hollow-cylindrical dielectric crystal of rutile is placed inside of an oxygen-free copper enclosure, as shown in Fig. \ref{fig:1_dielectricResonator}(a), with optical apertures and spaces both above and below to mount optical components.
Rutile's high dielectric constant ($\epsilon_\perp\approx130$ and $\epsilon_\parallel\approx255$ at 4K \cite{sabisky_measurements_2004, tobar_anisotropic_1998, klein_dielectric_1995}) makes the microwave electric field confined mostly inside the crystal [Fig.\ref{fig:1_dielectricResonator}(b)]. 
As a result, the dissipation on the enclosure wall becomes negligible, and the rutile's low loss tangent at microwave frequencies at low temperatures\cite{sabisky_measurements_2004,tobar_anisotropic_1998,klein_dielectric_1995} enables the internal quality factor to potentially reach $\sim\,10^6$, comparable to the state-of-the-art planar superconducting resonators.\cite{megrant_planar_2012, bruno_reducing_2015, calusine_analysis_2018}
The resonator's vacuum magnetic field of 5.0 pT yields a coupling constant of about $g_{\mathrm{single}}/2\pi = 0.07\,\mathrm{Hz}$ to a \textit{single} radical spin in DPPH or a \textit{single} P1 center (supplementary material). 
This value is comparable to those reported in other three-dimensional resonators.\cite{ball_loop-gap_2018, kato_high-cooperativity_2023}
The resonator's $\mathrm{TE_{01\delta}}$ mode (fundamental mode), which is of interest, is designed to have a resonance frequency of around 5 to 6 GHz to match up with the typical microwave photon frequencies in superconducting quantum circuits. 

The dielectric resonator is inductively coupled to the measurement line through a loop antenna, as depicted in Fig. \ref{fig:1_dielectricResonator}(b)], and positioned at the center of a superconducting three-dimensional vector magnet. 
The device is thermalized to the mixing chamber stage in a dilution refrigerator. 
The superconducting magnet generates an external constant magnetic field $\mathbf{B}_{\mathrm{DC}}$ oriented perpendicular to the microwave magnetic field $\mathbf{B}_{\mathrm{AC}}$, as illustrated in Fig. \ref{fig:1_dielectricResonator}(b). 
Two distinct microwave input lines with different total attenuation are fed to the mixing chamber stage. 
The heavily attenuated line is dedicated to continuous wave electron spin resonance (cw-ESR) spectroscopy, whereas the other less attenuated line supports pulse electron spin resonance (pulse-ESR) spectroscopy, which generally requires large microwave power for rapid spin driving.
The reflected signal is routed to the output line through a series of circulators and amplifiers and finally detected by a vector network analyzer (VNA) or a digitizer. 
For further details on the experimental setup and wiring, see the supplementary material.

Figure \ref{fig:1_dielectricResonator}(c) shows the reflection spectrum of the dielectric resonator measured at 10 mK, revealing the resonance frequency $\omega_r/2\pi$ of approximately 5.534 GHz. 
The internal quality factor $Q_{\mathrm{int}}$, measured to be $2.30\times10^4$, is predominantly limited by radiation losses through the optical apertures, as detailed in the supplementary material. 
Nevertheless, this $Q_{\mathrm{int}}$ already surpasses by more than an order of magnitude the values achieved in other typical three-dimensional copper resonator configurations designed for magnetic coupling to spin ensembles, such as loop-gap resonators \cite{fernandez-gonzalvo_cavity-enhanced_2019,ball_loop-gap_2018} or re-entrant cavities.\cite{goryachev_high-cooperativity_2014, menke_reconfigurable_2017}
Indeed, this value is consistent with the results from a similar dielectric resonator coupled to a magnon mode in a yttrium-ion-garnet (YIG) crystal at 4 K.\cite{kato_high-cooperativity_2023}

\begin{figure}
\includegraphics[width=\hsize]{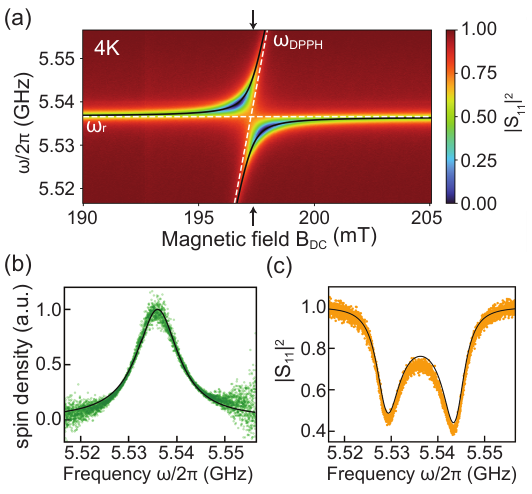}
\caption{\label{fig:2_DPPHcwESR} Spectroscopy of DPPH at 4 K.  (a) Color plot of the resonator reflection spectra $|S_{11} (\omega)|$ with white dotted lines indicating the original eigenfrequencies of DPPH ($\omega_{\mathrm{DPPH}}$) and the resonator ($\omega_r$). 
Black solid lines represent the dressed eigenfrequencies of the spin-resonator system. 
(b) Spin distribution estimated from the VNA spectrum at $197.25\,\mathrm{mT}$, where green dots and black solid line represent the estimated spin population and the Lorentzian fitting.
(c) Fitting results of the reflection spectrum at $\mathbf{B}_{\mathrm{DC}}=197.25\,\mathrm{mT}$, with 
yellow dots and the black line indicating the measured data and double Lorentzian fitting, respectively.}
\end{figure}

First, we tested with the crystalline powder of 2,2-diphenyl-1-picrylhydrazyl (DPPH), which possesses a free radial ($S=1/2$) per molecule and is commonly used as a standard sample in ESR studies. 
All measurements on DPPH were conducted at $4\,\mathrm{K}$. We skipped measuring below this temperature as DPPH goes through magnetic phase transitions from a paramagnetic to an antiferromagnetic phase,\cite{mergenthaler_strong_2017} leading to complex spectra that would complicate the analysis. 

The reflection spectra of the dielectric resonator containing DPPH are color-plotted versus $\mathbf{B}_{\mathrm{DC}}$ in Fig. \ref{fig:2_DPPHcwESR} (a). 
The avoided level crossing is observed when the DPPH's nominal transition frequency approaches the resonator frequency.
The black curve in Fig. \ref{fig:2_DPPHcwESR}(a) are derived through fitting with the ensemble coupling constant $g_\mathrm{ens,\mathrm{DPPH}}$ as a free parameter. 
The total linewidth of the resonator, $\kappa_\mathrm{tot}$ was estimated from VNA spectrum at $\mathbf{B}_\mathrm{DC}=0\,\mathrm{mT}$, and then the inhomogeneous broadening of the DPPH spins ($\Gamma_{\mathrm{DPPH}}$) was estimated from the spin density distribution [Fig.\ref{fig:2_DPPHcwESR} (b)]. This distribution was obtained by deducting the bare resonator spectrum at $\mathbf{B}_{\mathrm{DC}}=197.25\ \mathrm{mT}$ through the input-output theories.\cite{grezes_multimode_2014}
The fittings yield $g_{\mathrm{ens,DPPH}}/2\pi=7.8\ \mathrm{MHz}$, $\kappa_\mathrm{tot}/2\pi=1.7\,\mathrm{MHz}$, $\Gamma_{\mathrm{DPPH}}/2\pi = 9.6\ \mathrm{MHz}$, leading to a cooperativity $ C=g_\mathrm{ens}^2/(\kappa\Gamma)=3.73$, manifesting the system being in the high-cooperativity regime.  
Figure \ref{fig:2_DPPHcwESR} (c) shows the spectrum at $197.25\,\mathrm{mT}$ [indicated by white arrows in Fig. \ref{fig:2_DPPHcwESR} (a)], alongside a theoretical curve with above parameters. 

The inhomogeneous linewidth, approximately $10\,\mathrm{MHz}$, observed in our sample is narrower than the $15\,\mathrm{MHz}$ linewidth measured in a single crystalline DPPH with a higher spin concentration at $4\,\mathrm{K}$.\cite{mergenthaler_strong_2017}
However, it is twice as wide as the linewidth typically observed at room temperatures.\cite{abe_electron_2011}
The reason behind is likely due to its transition into an antiferromagnetic phase, but which is beyond the scope of this work. 

\begin{figure}
\includegraphics[width=\hsize]{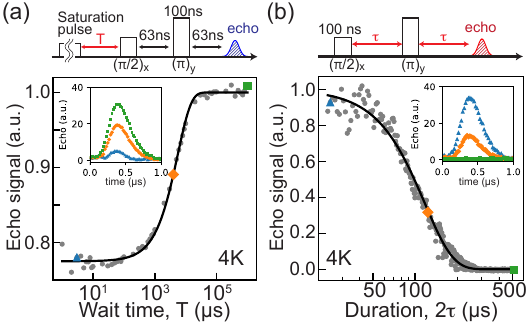}
\caption{\label{fig:3_DPPHpulseESR} Pulse-ESR of DPPH at $4\,\mathrm{K}$. 
We set 100 ms of cooldown time between each pulse sequence.
(a) Saturation recovery measurement results. 
The longitudinal relaxation time is measured to be $T_1 = 5.54 \pm 0.17\, \mathrm{ms}$. 
The upper panel illustrates the pulse sequence, while the inset shows time traces of echoes measured at different time intervals, $T = 16\,\mathrm{\mu s}, 4\,\mathrm{ms}$, and $1\,\mathrm{s}$, marked by blue triangles, orange diamonds, and green squares, respectively. 
(b) Hahn-echo measurement results. 
The transverse relaxation time is measured to be $T_2 = (1.173 \pm 0.008) \times 10^2\,  \mathrm{\mu s}$. 
The upper panel illustrates the pulse sequence, while the inset shows time traces of echoes measured at different time intervals ($2\mathrm{\tau} = 38, 137, 527\, \mathrm{\mu s}$), marked by blue triangles, orange diamonds, and green squares. 
}
\end{figure}

We also demonstrate pulse-ESR spectroscopy on DPPH using two pulse sequences: saturation recovery and Hahn-echo, as depicted in the upper panels of Fig. \ref{fig:3_DPPHpulseESR} (a) and (b). 
The longitudinal and the transverse relaxation times are measured to be $T_1 = 5.54 \pm 0.17 \, \mathrm{ms}$ and $T_2 = (1.173 \pm 0.008) \times 10^2 \, \mu \mathrm{s}$, respectively. 
For the $T_2$ measurement analysis, we applied the stretched exponential function, $\exp{[-(\frac{2\tau}{T_2})^p]}$, yielding $p=2.1$. 
While relaxation times for DPPH at 4K have not been previously reported, our findings show no contradiction with the values measured at $20\,\mathrm{K}$,\cite{meyer_temperature_2013} taking into account much denser spin concentration in our DPPH sample. 
The observed gaussian-like decay in $T_2$ indicates a magnetically fluctuating environment, likely due to the nuclear spins of $^{14}\mathrm{N}$ and $^{13}\mathrm{C}$, as well as possible fluctuations from antiferromagnetic spins.

\begin{figure}
    \includegraphics[width=\hsize]{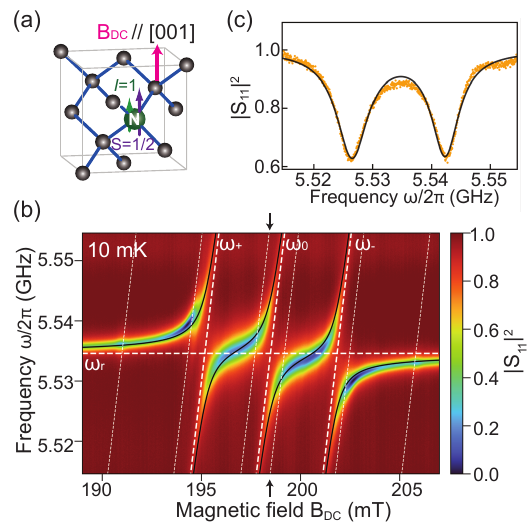}
    \caption{\label{fig:4_P1center} Spectroscopy of P1 centers in diamond. 
    (a) Schematic of a P1 center in a diamond crystal, consisting of electron spin ($S=1/2$) and a nuclear spin ($I=1$) of $\mathrm{^{14}N}$. 
    The static magnetic field $B_\mathrm{{DC}}$ is aligned along the $[001]$ direction. 
    (b) Color plot of the resonator reflection spectra $|S_{11} (\omega)|$, with white dashed lines indicating the original eigenfrequencies of each nuclear spin state ($\omega_+,\omega_0,\omega_-$) and the dielectric resonator ($\omega_r$). 
    The transitions of P1 centers hyperfine-coupled to $\mathrm{^{13}C}$ nuclei are marked with white dashed lines. 
    (c) The spectrum at 198.48 mT, with yellow dots and black curve indicating experimental data and double Lorentzian fitting, respectively (see supplementary material for more details).}
\end{figure}

Subsequently, we measured the resonator with a diamond crystal, intended for use as a quantum transducer, containing substitutional nitrogen impurities, known as P1 centers [Fig. \ref{fig:4_P1center}(a)]. 
The estimated concentration of P1 centers for this sample is $106\,\mathrm{ppm}$ (supplementary material). 
The spin Hamiltonian of P1 center consisting of an electron spin $S=1/2$ and a nuclear spin $I=1$ of $^{14}\mathrm{N}$ is the following 
\begin{equation}
H_{P1}/\hbar = \gamma_e \mathbf{B_{\mathrm{DC}}}\mathbf{S}  + \mathbf{S} \mathcal{A}_{\mathrm{N'}} \mathbf{I}, 
\end{equation}
where $\mathbf{S}$ and $\mathbf{I}$ are the spin operators of the electron and of $^{14}\mathrm{N}$ nucleus, and $\mathcal{A}_{\mathrm{N'}}\, (\mathcal{A}_{\mathrm{N'}\perp}/2\pi = 114.03\,    \mathrm{MHz}, \mathcal{A}_{\mathrm{N'}\parallel}/2\pi = 81.33\,    \mathrm{MHz})$ is the hyperfine tensor for the $^{14}\mathrm{N}$ nuclear spin. 
Figure \ref{fig:4_P1center}(b) reveals three avoided level crossings in the resonator's spectroscopy versus $\mathbf{B}_{\mathrm{DC}} \| [001]$, where all the four possible orientations of P1 centers are degenerated, at 10 mK.
The three avoided level crossings arise from the nuclear-spin-conserving ESR transitions: $\omega_{+}, \omega_{0}$, and $\omega_{-}$, associated with the three nuclear spin states and displayed by white dashed lines.
The ensemble coupling constants are estimated to be $g_\mathrm{ens,+}/2\pi=9.2\ \mathrm{MHz}$, $g_\mathrm{ens,0}/2\pi=9.3\ \mathrm{MHz}$, and $g_\mathrm{ens,-}/2\pi=8.5\ \mathrm{MHz}$ (supplementary material), where the symbol $i$ in $g_\mathrm{ens,i}$ corresponds to the nuclear spin state. In addition, $\mathrm{^{13}C}$ hyperfine-coupled P1 centers were measured and indicated by the thin white dashed lines in Fig. \ref{fig:4_P1center}(b).
The reflection spectrum at $198.48 \mathrm{mT}$, which shows an avoided level crossing of $I=0$ transition, is plotted in Fig. \ref{fig:4_P1center}(c).
Similar to DPPH, we determined the inhomogeneous broadening of the P1 centers, $\Gamma_{P1}/2\pi\approx1.0\ \mathrm{MHz}$, from the VNA spectra, as detailed in the supplementary material. 
Accordingly, based on the estimated parameters, we find the spins are strongly coupled to the resonator ($g_\mathrm{ens}\gg\kappa_\mathrm{tot},\Gamma_{\mathrm{P1}}$).
It is important to note that the linewidth of the bare resonator slightly changed to $\kappa_\mathrm{tot}/2\pi\approx1.9 \mathrm{MHz}$ due to a slight shift in the position of the coupling loop-antenna. 


\begin{figure}
    \includegraphics[width=\hsize]{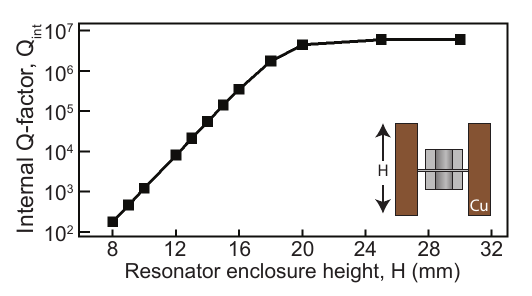}

    \caption{\label{fig:Qint} The simulated internal quality factor versus resonator enclosure height, $H$. 
    The rutile and copper enclosure dimensions are unchanged except for the height. 
    Consider that the present configuration with a height of $H=10\,\mathrm{mm}$ exhibits a tenfold increase in $Q_{int}$ compared to this simulation result. This enhancement arises from the extension of an effective resonator enclosure height, including spaces for optics both above and below.
    }
\end{figure}

Lastly,  we explore methods to enhance the internal quality factor $Q_{\mathrm{int}}$ through numerical simulations as shown in Fig. \ref{fig:Qint}. 
The results indicate a substantial reduction in radiation loss by increasing the resonator enclosure height $H$. When $ H \ge 18~\mathrm{mm}$, $Q_{\mathrm{int}}$ exceeds $10^6$, primarily limited by the microwave loss tangent of rutile.\cite{klein_dielectric_1995} Further details on the numerical simulation conditions can be found in the supplementary material.
Importantly, both the resonance frequency and single spin coupling vary insignificantly within $0.8\% $ and $10.8\%$, respectively, across a range of resonator enclosure heights from $8.0\mathrm{mm}$ to $30.0 \mathrm{mm}$. 
These results also emphasize that the enclosure height has a minimal impact on the resonator mode volume, as the microwave electric fields are predominantly confined inside the rutile.

In conclusion, we demonstrated a microwave dielectric resonator with an internal quality factor of $Q_{\mathrm{int}} \sim 10^4$ with 8 mm optical apertures. 
The internal quality factor can even be enhanced to above $10^6$ by simple modification of the copper enclosure. 
We performed both continuous-wave and pulse electron spin resonance spectroscopies on a DPPH powder and a diamond crystal. 
Once the internal quality factor exceeds $10^5$, the resonator's quality factor at the single microwave photon power level, where microscopic defects would be the dominant source of the loss tangent, can be investigated. It is noteworthy that no significant microwave input power dependence of the quality factor has been observed with the current resonator.

\section*{ACKNOWLEDGMENTS}
The authors acknowledge K. Kato and K. Usami for sharing insights on rutile.
The authors also thank E. Abe and members of the Hybrid Quantum Device Team within the Science and Technology Group, as well as the Experimental Quantum Information Physics Unit, for discussion. 
This work has been supported by the JST Moonshot R\&D Program (Grant No. JPMJMS2066), JST-PRESTO (Grant No. JPMJPR15P7), The Nakajima Foundation, and The Sumitomo Foundation.

\section*{AUTHOR DECLARATIONS}
\subsection*{Conflict of Interest}
The authors have no conflicts to disclose.

\section*{Author Contributions}
Y.K. conceived and supervised the project alongside H.T.
R.K.B., T.H., and Y.K. initiated the simulation and design of the resonator, with A.B. later making significant modifications. T.H., A.B., and Y.K. performed the measurement and analysis. 
All authors contributed to writing the manuscript. 

\section*{DATA AVAILABILITY} 
The data supporting this study's findings are available from the corresponding author upon reasonable request.


\bibliography{references}

\end{document}